\documentclass[conference]{IEEEtran}
\IEEEoverridecommandlockouts
\usepackage{url}
\usepackage{cite}
\usepackage{amsmath,amssymb,amsfonts}
\usepackage{algorithmic}
\usepackage{graphicx}
\usepackage{textcomp}
\usepackage{xcolor}
\usepackage{hyperref}
\usepackage{orcidlink}
\def\BibTeX{{\rm B\kern-.05em{\sc i\kern-.025em b}\kern-.08em
    T\kern-.1667em\lower.7ex\hbox{E}\kern-.125emX}}
\begin{document}

\title{Hybrid Deep Learning and Handcrafted Feature Fusion for Mammographic Breast Cancer Classification\\
\thanks{}
}

\author{
\IEEEauthorblockN{Maximilian Tschuchnig\,\orcidlink{0000-0002-1441-4752}\IEEEauthorrefmark{1}\IEEEauthorrefmark{2}\IEEEauthorrefmark{3},
Michael Gadermayr\,\orcidlink{0000-0003-1450-9222}\IEEEauthorrefmark{1},
Khalifa Djemal\IEEEauthorrefmark{3}}

\IEEEauthorblockA{\IEEEauthorrefmark{1} Information Technologies and Digitalisation, Salzburg University of Applied Sciences, Austria}
\IEEEauthorblockA{\IEEEauthorrefmark{2}Artificial Intelligence and Human Interfaces, University of Salzburg, Austria}
\IEEEauthorblockA{\IEEEauthorrefmark{3}Laboratoire IBISC, University of Evry Paris-Saclay, France}

\IEEEauthorblockA{maximilian.tschuchnig@fh-salzburg.ac.at}
}

\maketitle

\begin{abstract}
Automated breast cancer classification from mammography remains a significant challenge due to subtle distinctions between benign and malignant tissue. In this work, we present a hybrid framework combining deep convolutional features from a ResNet-50 backbone with handcrafted descriptors and transformer-based embeddings. Using the CBIS-DDSM dataset, we benchmark our ResNet-50 baseline (AUC: $78.1\%$) and demonstrate that fusing handcrafted features with deep ResNet-50 and DINOv2 features improves AUC to $79.6\%$ (setup \textit{d1}), with a peak recall of $80.5\%$ (setup \textit{d1}) and highest F1 score of $67.4\%$ (setup \textit{d1}). Our experiments show that handcrafted features not only complement deep representations but also enhance performance beyond transformer-based embeddings. This hybrid fusion approach achieves results comparable to state-of-the-art methods while maintaining architectural simplicity and computational efficiency, making it a practical and effective solution for clinical decision support.
\end{abstract}


\section{Introduction}
Breast cancer screening via mammography may reduce mortality through early detection, as suggested by observational evidence from modern screening programs~\cite{weedon2014modern}. However, radiologist workload~\cite{hoff2019influence} and diagnostic variability~\cite{elmore2009variability} remain significant challenges. For instance, when radiologists exceed $10{,}000$ mammograms per year, this has been associated with reduced sensitivity and increased false-positive rates~\cite{hoff2019influence}.

Early computer-aided detection (CAD) systems attempted to address these issues using hand-engineered features derived from radiological knowledge and classical pattern recognition. While these traditional CAD approaches demonstrated potential, they often failed to generalize effectively and frequently increased recall rates without meaningful gains in diagnostic accuracy. The rise of deep learning (DL), particularly convolutional neural networks (CNNs), marked the turning point for data-driven architectures to outperform traditional methods by learning hierarchical feature representations directly from imaging data.

While these deep learning methods have demonstrated remarkable efficacy across domains, pure end-to-end architectures, particularly those pretrained on natural image datasets like ImageNet~\cite{deng2009imagenet}, often require fine-tuning to effectively model domain-specific features in medical imaging~\cite{usman2022analyzing}. This becomes especially problematic when training data are scarce or when domain shifts occur between the source and target distributions~\cite{anton2022well}. To address these limitations, recent work has highlighted the utility of hybrid learning frameworks that fuse deep, learned features with handcrafted descriptors~\cite{foleis2025transfer}. Such models can leverage both global patterns and fine-grained, domain-specific cues—elements often underutilized by convolutional filters alone. Further adaptations to conventional deep learning methods employ path-based and multi-scale features, especially if the image data is dense in pixels~\cite{quintana2023exploiting,petrini2022breast}.

In parallel, transformer-based architectures such as the Vision Transformer (ViT)~\cite{dosovitskiy2020image} and its self-supervised variant DINOv2~\cite{oquab2023dinov2} have introduced fundamentally different paradigms for image representation. Unlike CNNs, ViTs use attention mechanisms to model long-range dependencies, enabling them to extract context-aware global features. These properties make transformer-based embeddings highly complementary to handcrafted features and CNN-derived features alike.

We propose a feature fusion framework for breast cancer classification that enhances a ResNet-50~\cite{he2016deep} backbone with handcrafted and ViT descriptors. Our method improves baseline AUC, F1-score, and recall. Comprehensive ablation studies show that the integration of handcrafted features boosts AUC by $2.8\%$ and F1-score by $2.9\%$, demonstrating the complementary value of traditional descriptors and attention-based representations in modern medical image analysis.

\section{Methodology}
\begin{figure*}[ht]
    \centering
    \includegraphics[width=0.8\linewidth]{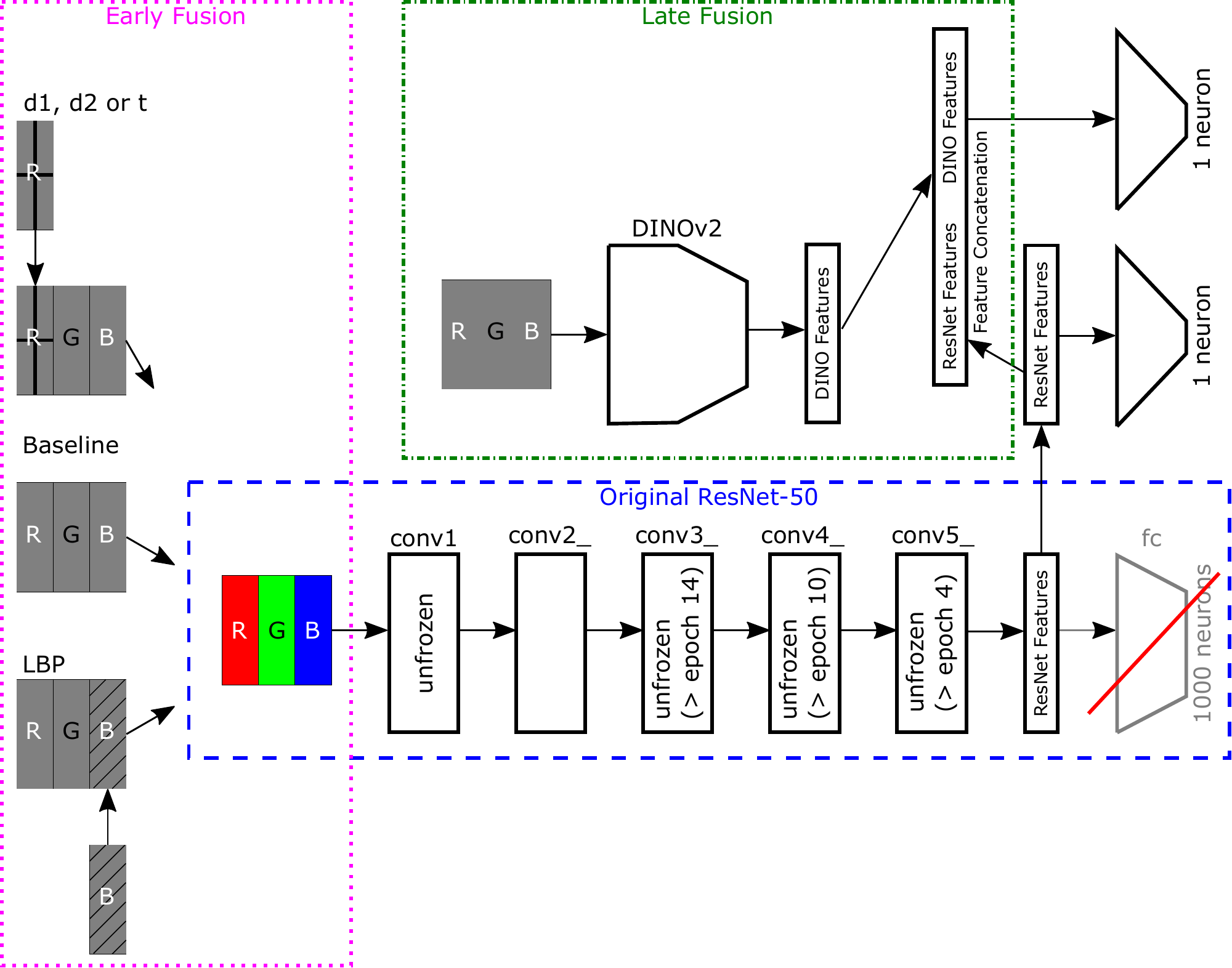}
    \caption{Overview of the proposed hybrid deep learning setups, showing the baseline and all ablated fusion configurations.}
    \label{fig:big_picture}
\end{figure*}

To establish an effective, deep, multifeature model we need a combination of a relevant dataset, features and a baseline to adapt. To this end, we use the Curated Breast Imaging Subset of DDSM (CBIS-DDSM)~\cite{lee2017curated} dataset, a mixture of handcrafted features and Dinov2 features~\cite{oquab2023dinov2} and finally ResNet-50, as our baseline model. We apply transfer learning to convert the imagenet pretrained ResNet-50 to the medical image domain and use an ablation study to investigate the effect of the different features on the performance measures Area-under-the-Curve (AUC), Accuracy (ACC), Precision (PRE), Recall (Rec) and F1-Score (F1). The complete methodology is illustrated in Fig.~\ref{fig:big_picture}, which details the baseline architecture, the integration of handcrafted features through early fusion, and the incorporation of DINO features via late fusion.

\subsection{Dataset}
The CBIS-DDSM~\cite{lee2017curated} is a publicly available dataset of digitized screening mammograms, annotated by expert radiologists with regions of interest (ROIs) corresponding to suspicious findings. For this study, we utilized the calcification and mass case subsets available via Kaggle (\url{https://www.kaggle.com/datasets/awsaf49/cbis-ddsm-breast-cancer-image-dataset}). From this resource, we extracted the cropped ROI images along with their corresponding pathology labels. Our training set consisted of $1,328$ benign and $1,428$ malignant ROI images, while the held-out test set included $428$ benign and $276$ malignant samples. The training set was further split into training ($80\%$) and validation ($20\%$) subsets. This dataset offers a clinically relevant benchmark with high-quality, expert-labeled annotations, making it well suited for assessing the effectiveness and generalizability of our proposed feature fusion framework.

For data loading and augmentation, four separate data pipeline configurations were defined, one for training and another shared between validation and testing. All images were initially resized to $600 \times 600$ pixels for training, and to $512 \times 512$ pixels for validation and testing. Each image was then converted into a PyTorch tensor and normalized using ImageNet statistics to maintain compatibility with the pretrained backbone. To enhance variability and robustness, the training pipeline additionally applied random resized cropping to $512 \times 512$ pixels with a scaling factor sampled uniformly from the range $(0.9, 1.0)$. Additionally, horizontal flipping was applied with a probability of $50\%$ to introduce orientation invariance, as CBIS-DDSM mammographic images do not have a fixed horizontal alignment.

\subsection{Feature Extraction}
In the baseline configuration, grayscale CBIS-DDSM images are converted to RGB format by duplicating the single grayscale channel across all three RGB channels. This conversion facilitates compatibility with the ResNet-50 architecture and enables straightforward integration of additional feature channels.

To enrich the input with handcrafted features, we explore the use of horizontal and vertical edge detection (\textit{d1}), second-order derivatives (\textit{d2}), simple thresholding (\textit{t}), and Local Binary Patterns (\textit{LBP}). These features are applied individually and in various combinations. While \textit{d1} and \textit{d2} can, in theory, be learned by convolutional layers and may offer limited additional value, features like thresholding and LBP are more challenging for neural networks to infer and may contribute unique information.

When \textit{d1}, (\textit{d2}, or \textit{t} are selected, the corresponding feature map is computed, either via convolution (for \textit{d1} and \textit{d2}) or thresholding (for \textit{t}), and inserted into the red channel of the RGB image. This constitutes an early fusion strategy~\cite{zhang2021deep,zhang2021modality}, where both the original grayscale image and the derived feature are simultaneously provided to the network. If multiple features are used, their values are averaged before being placed in the red channel to preserve dynamic range. For \textit{LBP}, we compute a local binary pattern with an $8$-neighbor configuration over a $3 \times 3$ window. The resulting LBP map is inserted into the blue channel of the RGB image. This modular fusion approach allows handcrafted features to be embedded into standard network pipelines without modifying model architecture.

Transformer-based models such as the Vision Transformer (ViT)\cite{dosovitskiy2020image}, and in particular its self-supervised variant DINOv2\cite{oquab2023dinov2}, are employed as complementary deep learning-based feature extractors. Unlike CNNs, which rely on local convolutional filters and thus have an inherently limited receptive field, ViTs use multi-head self-attention mechanisms. This architecture enables ViTs to model long-range dependencies and capture global context across the entire image, potentially producing feature representations that differ intrinsically from those of convolutional networks.

To integrate transformer-based features, the grayscale input image is passed through the pretrained DINOv2-small model from Meta AI, yielding a $384$ dimensional embedding. These features are then concatenated with the output of the final ResNet-50 layer, implementing a late fusion strategy~\cite{zhang2021deep,zhang2021modality}. The combined representation is subsequently passed through a modified classification head to predict whether the input region of interest is benign or malignant.

\subsection{Deep, Multifeature Tumor Detection}
We adopt ResNet-50, pretrained on ImageNet, as the primary backbone and baseline architecture. To adapt the network for binary classification (malignant vs. benign), the original classification head is replaced with a custom head consisting of a fully connected layer with $256$ neurons, followed by a ReLU activation and a dropout layer with a dropout rate of $0.4$. This is followed by a final linear layer with a single output neuron, optimized using the Binary Cross Entropy (BCE) loss with logits.

In the baseline setup, all pretrained layers of ResNet-50 are frozen, and only the new classification head is fine-tuned. In subsequent experiments, we progressively unfreeze additional ResNet layers in stages to enable deeper fine-tuning. Specifically, we unfreeze the final stage at epoch $4$, stage $4$ at epoch $10$, and stage $3$ at epoch $14$. At each unfreezing step, the learning rate for the newly trainable parameters is scaled down by a factor of $0.1$ relative to the previous layers, helping to mitigate catastrophic forgetting and promote stable training. Furthermore, in all configurations except the baseline, the initial convolutional layer is also unfrozen to allow the network to better adapt to early-fused input signals. All parameters were tuned to optimize validation performance.

All models were trained for $25$ epochs using the ADAM optimizer with a batch size of $32$, a learning rate of $10^{-4}$, and a weight decay of $10^{-3}$. A cosine annealing learning rate scheduler was applied, with a step size of $10$ epochs and a gamma of $0.1$. Label smoothing with a factor of $0.1$ was used to promote smoother training. Each setup was evaluated twice under different random seeds to ensure stability and robustness of the results.

\section{Results}

\begin{figure}[ht]
    \centering
    \includegraphics[width=\linewidth]{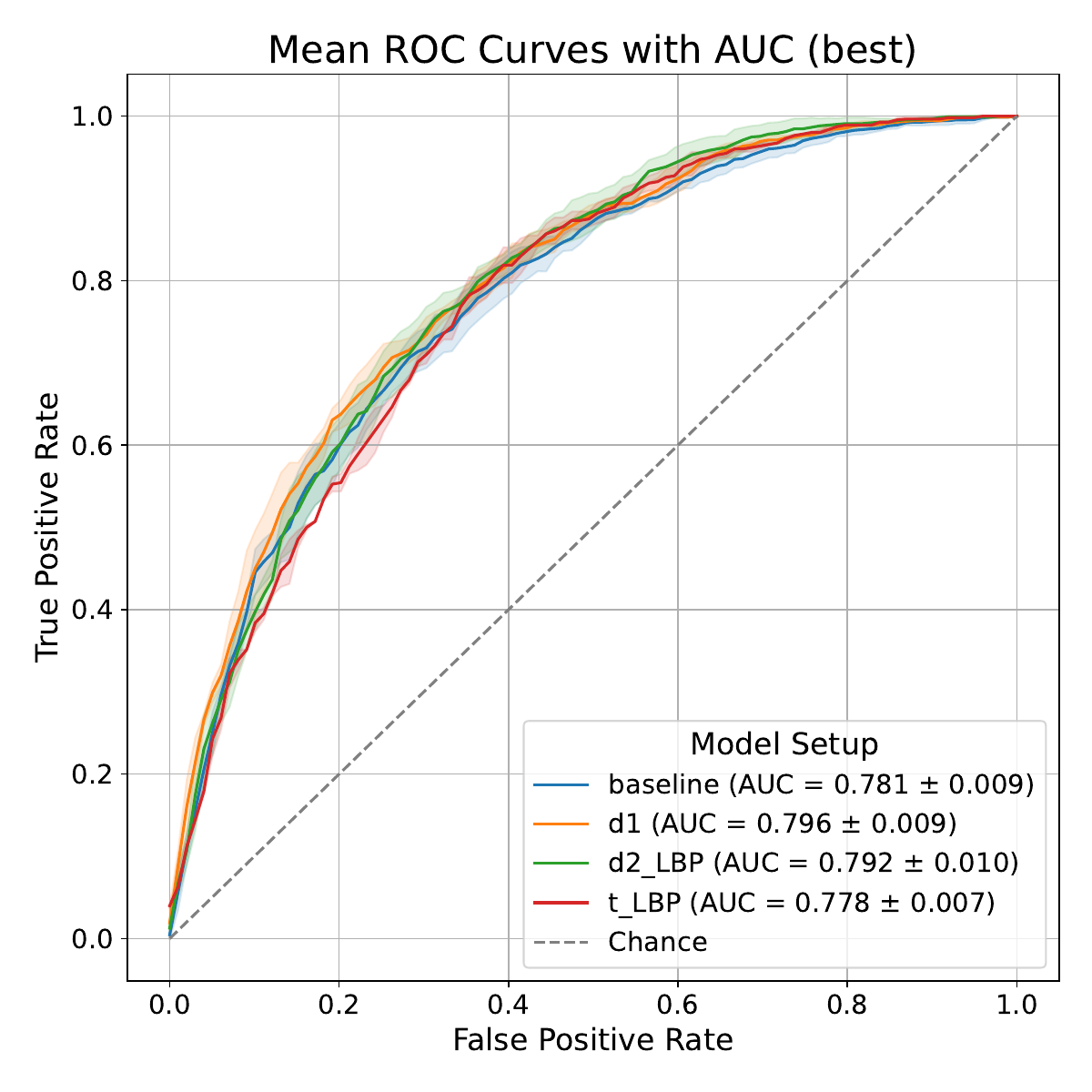}
    \caption{ROC curves of the top-3 configurations by AUC and F1-score, compared to the baseline. Plots show the mean over four runs, shaded regions denote $\pm1$ standard deviation.}
    \label{fig:best}
\end{figure}

\begin{figure}[ht]
    \centering
    \includegraphics[width=\linewidth]{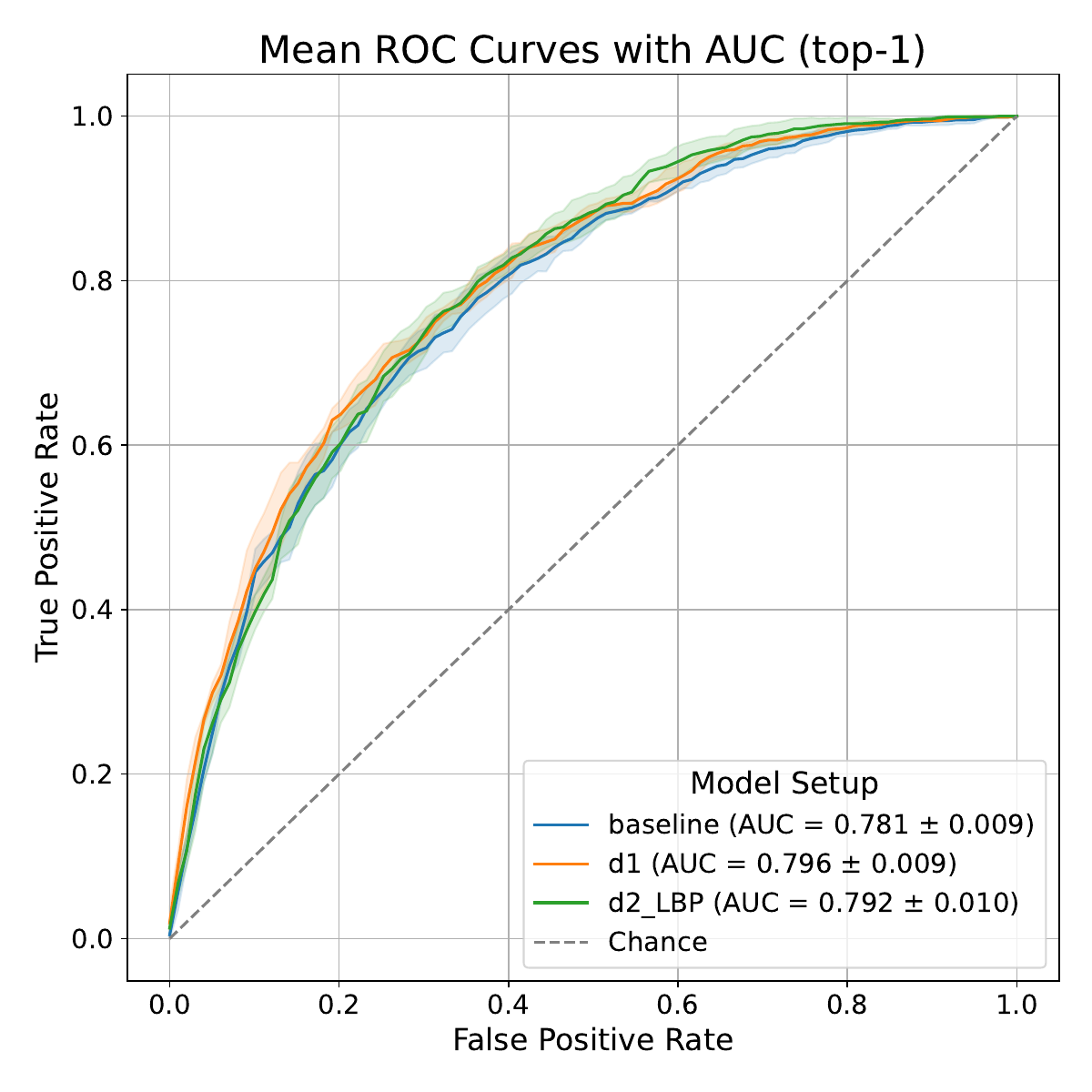}
    \caption{ROC curves of the top-1 configurations by AUC and F1-score, compared to the baseline. Plots show the mean over four runs, shaded regions denote $\pm1$ standard deviation.}
    \label{fig:vbest}
\end{figure}

Table~\ref{Tab:AUC} and Table~\ref{Tab:F1} report the averaged classification performance across four experimental runs. Table~\ref{Tab:AUC} is sorted in descending order of Area Under the ROC Curve (AUC), while Table~\ref{Tab:F1} is sorted by descending F1-score. The reported metrics include AUC, F1-score, precision (PRC), recall (REC), and accuracy (ACC), computed on the held-out test set. Across both rankings, the best performing individual setup in terms of AUC is the \textit{d1} configuration (AUC = $0.796$), followed closely by \textit{d2\_LBP} ($0.791$) and \textit{dino\_d2} (0.787). When sorted by F1-score, \textit{d1} yields the highest score again (F1 = $0.674$), followed by \textit{d2\_LBP} ($0.672$) and \textit{t\_LBP} ($0.672$). % All top-performing configurations show statistically significant improvements over the baseline, as determined by paired t-tests with a significance level of $\alpha = 0.05$. 
These configurations combine handcrafted filter-based representations (first- and second-order derivatives and local binary patterns) in various permutations, sometimes with extracted DINOv2 features. Fig.\ref{fig:best} shows the mean ROC curves of the top-3 configurations (ranked by AUC and F1-score), along with standard deviation as uncertainty bands, compared to the baseline. Fig.\ref{fig:vbest} illustrates the ROC curves of the two best-performing configurations individually with \textit{d1} (highest AUC) and \textit{d2\_LBP} (highest F1-score).

Several configurations that incorporate handcrafted features, either as standalone (e.g., \textit{d1}, \textit{d2}) or in fused representations (e.g., \textit{d2\_LBP}, \textit{d1\_LBP}, \textit{dino\_d2}), consistently outperform the baseline and often, \textit{dino} configurations on AUC and F1-score. In contrast, the \textit{frozen} model yields the lowest scores across all metrics, highlighting the limited effectiveness of unadapted pretrained imagenet representations in this setting.

\begin{table}[]
\centering
\caption{Mean results over four experimental runs. The table is sorted by descending mean AUC.}
\label{Tab:AUC}
\begin{tabular}{l|lllll}
setup             & AUC            & F1    & PRC   & REC   & ACC   \\ \hline
d1                & 0.796          & 0.674 & 0.581 & 0.805 & 0.694 \\
d2\_LBP           & 0.791          & 0.672 & 0.596 & 0.772 & 0.705 \\
dino\_d2 & 0.787 & 0.664 & 0.598 & 0.748 & 0.704 \\
d2                & 0.787          & 0.663 & 0.609 & 0.730 & 0.710 \\
d1\_LBP           & 0.786          & 0.661 & 0.585 & 0.764 & 0.692 \\
dino\_d1          & 0.782          & 0.671 & 0.591 & 0.779 & 0.702 \\
dino\_all         & 0.781          & 0.659 & 0.601 & 0.728 & 0.704 \\
baseline          & 0.781          & 0.659 & 0.594 & 0.741 & 0.699 \\
t\_LBP            & 0.778          & 0.672 & 0.583 & 0.793 & 0.697 \\
LBP               & 0.776          & 0.658 & 0.592 & 0.740 & 0.698 \\
dino\_LBP         & 0.774          & 0.646 & 0.591 & 0.712 & 0.694 \\
all               & 0.774          & 0.656 & 0.575 & 0.764 & 0.686 \\
t                 & 0.769          & 0.653 & 0.581 & 0.746 & 0.689 \\
dino\_t           & 0.765          & 0.647 & 0.575 & 0.743 & 0.682 \\
dino              & 0.759          & 0.656 & 0.579 & 0.757 & 0.688 \\
frozen            & 0.693          & 0.588 & 0.517 & 0.683 & 0.626
\end{tabular}
\end{table}

\begin{table}[]
\centering
\caption{Mean results over four experimental runs. The table is sorted by descending mean F1.}
\label{Tab:F1}
\begin{tabular}{l|lllll}
setup     & AUC   & F1    & PRC   & REC   & ACC   \\ \hline
d1        & 0.796 & 0.674 & 0.581 & 0.805 & 0.694 \\
d2\_LBP   & 0.791 & 0.672 & 0.596 & 0.772 & 0.705 \\
t\_LBP    & 0.778 & 0.672 & 0.583 & 0.793 & 0.697 \\
dino\_d1  & 0.782 & 0.671 & 0.591 & 0.779 & 0.702 \\
dino\_d2  & 0.787 & 0.664 & 0.598 & 0.748 & 0.704 \\
d2        & 0.787 & 0.663 & 0.609 & 0.730 & 0.710 \\
d1\_LBP   & 0.786 & 0.661 & 0.585 & 0.764 & 0.692 \\
baseline  & 0.781 & 0.659 & 0.594 & 0.741 & 0.699 \\
dino\_all & 0.781 & 0.659 & 0.601 & 0.728 & 0.704 \\
LBP       & 0.776 & 0.658 & 0.592 & 0.740 & 0.698 \\
all       & 0.774 & 0.656 & 0.575 & 0.764 & 0.686 \\
dino      & 0.759 & 0.656 & 0.579 & 0.757 & 0.688 \\
t         & 0.769 & 0.653 & 0.581 & 0.746 & 0.689 \\
dino\_t   & 0.765 & 0.647 & 0.575 & 0.743 & 0.682 \\
dino\_LBP & 0.774 & 0.646 & 0.591 & 0.712 & 0.694 \\
frozen    & 0.693 & 0.588 & 0.517 & 0.683 & 0.626
\end{tabular}
\end{table}

Furthermore, Table~\ref{Tab:comp} compares our best AUC with the ones reported on paperswithcode. Here, the proposed method stands out as not using any sort of additional patch based or multiscale processing, while reporting $1.3\%$ less AUC then first place and $0.7\%$ less than second place.

\begin{table}[]
\centering
\caption{Comparison of our reached AUC compared to leading CBIS-DDSM literature. Compared to all scores reported on paperswithcode:\url{https://paperswithcode.com/sota/cancer-no-cancer-per-image-classification-on}, our method reached a tied third place.}
\label{Tab:comp}
\begin{tabular}{l|l}
Setup                                   & AUC            \\ \hline
Multi-patch size DenseNet-121~\cite{quintana2023exploiting} & $0.809$ \\
SingleView\_PatchBased\_EfficientNet-B0~\cite{petrini2022breast} & $0.803$ \\
\textbf{Proposed Method (Average)} & \textbf{$0.796$} \\
MorphHR-ResNet18\_S896~\cite{wei2022beyond} & $0.796$ \\
ResNet18\_S896~\cite{wei2022beyond} & $0.796$ \\
SingleView\_PatchBased\_EfficientNet-B3~\cite{petrini2022breast} & $0.795$ \\
Multi-resolution DenseNet-121~\cite{quintana2023exploiting} & $0.789$ \\
ResNet18\_S448~\cite{wei2022beyond} & $0.788$ \\
Further results on paperswithcode: CBIS-DDSM & $< 785$      
\end{tabular}
\end{table}

\section{Discussion}
The results demonstrate that even a simple combination handcrafted filters and local texture descriptors leads to robust and competitive classification performance in mammographic image analysis. Specifically, the \textit{d1}, \textit{d2\_LBP}, \textit{dino\_d2} and \textit{t\_LBP} setups achieved the highest scores in AUC and F1, %significantly 
outperforming all purely deep learning-based feature extractors even in late fusion setups and most late fusion, transformer-based, \textit{dino} configurations.

Interestingly, these results are also highly competitive to highly sophisticated multi-resolution and (multi) patched approaches, beating most of them in terms of AUC and falling only $1.3\%$ behind the current first place on the leaderboard~\cite{quintana2023exploiting}, although using very simple early fusion setups.

This outcome supports the hypothesis that early fusion of domain-informed, handcrafted features with CNN-based architectures can be more effective than relying solely on pretrained representations. The performance gains observed with e.g. \textit{d1} or \textit{d2\_LBP} suggest that first- and second-order gradients, when used alone or combined with local binary patterns, enhance structural and microtextural representations crucial for distinguishing malignant from benign patterns in mammograms. Additional ROC and uncertainty investigations show that especially \textit{d2} is robust to random seeding, showing the importance of these additional very simple, features.

Moreover, the results show that sophisticated transformer-derived features can improve performance to the baseline but do not inherently outperform simpler, handcrafted representations when fused early in the input space. This observation reinforces prior evidence that CNNs often underutilize high-frequency or textural information, and that pretrained models, especially those trained on natural images like ImageNet, can struggle to generalize effectively to medical imaging domains without further adaptation.

These results contribute to the field by demonstrating that classical image processing techniques, when thoughtfully integrated, can not only complement but sometimes outperform modern self-supervised or transfer learning approaches. This finding is particularly relevant in medical imaging domains where training data is limited and domain-specific priors are essential. Overall, our study highlights the potential of hybrid input-level representations and motivates further research into lightweight, interpretable, and data-efficient approaches to augment deep learning based medical image classification tasks.

\section{Conclusion}
This study demonstrates that integrating handcrafted texture and structural features into a deep learning pipeline can %significantly
improve breast cancer classification performance in mammography. Our hybrid feature fusion approach, combining ResNet-50 with simple, predefined filters, thresholding and LBP, consistently outperform the baseline. Furthermore, while the addition of vision transformer based features did improve upon the baseline, this improvement was inferior to the addition of the simpler, handcrafted features. The results underscore the complementary nature of traditional image processing techniques and modern deep architectures, offering a reliable and computationally efficient framework for computer-aided diagnosis. Future work may explore dynamic fusion strategies or unsupervised feature selection to further enhance robustness and generalization.

\bibliographystyle{IEEEtran}
\bibliography{bibliography}

\section*{Acknowledgment}
This project was partly funded by the county of Salzburg under the project AIBIA, the Salzburg University of Applied Sciences under the project FHS-trampoline-8 (Applied Data Science Lab), and by ERASMUS+ through a staff mobility grant for training between FH Salzburg and Université d'Évry Val d'Essonne.

\end{document}